\documentclass[useAMS,usenatbib]{mnras}

\usepackage{times}
\usepackage{ amssymb }
\usepackage{mathptmx}
\usepackage{multirow}
\usepackage{psfrag}
\usepackage{pspicture}
\usepackage{graphicx}
\usepackage{amsmath}
\usepackage{bm}
\usepackage{color}
\usepackage{url}
\usepackage{caption}
\usepackage{subcaption}
\usepackage{grffile}
\usepackage[dvipsnames]{xcolor}
\usepackage{ulem}

\title[Convergence of pre-ICs]{Numerical convergence of pre-initial conditions on dark matter halo properties}

\author[Zhang et al.]
{Tianchi Zhang,$^{1, 2}$\thanks{Email: tczhang@bjp.org.cn}
Shihong Liao,$^{2, 3}$ Ming Li,$^{2}$ Jiajun Zhang$^{4}$\\
$^1$Beijing Planetarium, Beijing Academy of Science and Technology, Beijing 100044, China\\
$^2$Key Laboratory for Computational Astrophysics, National Astronomical Observatories, Chinese Academy of Sciences, Beijing 100012, China\\
$^3$Department of Physics, University of Helsinki, Gustaf Hällströmin katu 2, FI-00014 Helsinki, Finland\\
$^4$Center for Theoretical Physics of the Universe, Institute for Basic Science (IBS), Daejeon 34126, Korea\\
}
\begin{document} 



\maketitle

\label{firstpage}

\begin{abstract}
Generating pre-initial conditions (or particle loads) is the very first step to set up a cosmological $N$-body simulation. In this work, we revisit the numerical convergence of pre-initial conditions on dark matter halo properties using a set of simulations which only differs in initial particle loads, i.e. grid, glass, and the newly introduced capacity constrained Voronoi tessellation (CCVT). 
We find that the median halo properties agree fairly well (i.e. within a convergence level of a few per cent) among simulations running from different initial loads.
We also notice that for some individual haloes cross-matched among different simulations, the relative difference of their properties sometimes can be several tens of per cent.
By looking at the evolution history of these poorly converged haloes, we find that they are usually merging haloes or haloes have experienced recent merger events, and their merging processes in different simulations are out-of-sync, making the convergence of halo properties become poor temporarily. We show that, comparing to the simulation starting with an anisotropic grid load, the simulation with an isotropic CCVT load converges slightly better to the simulation with a glass load, which is also isotropic. 
Among simulations with different pre-initial conditions, haloes in higher density environments tend to have their properties converged slightly better.
Our results confirm that CCVT loads 
behave as well as the widely used grid and glass loads at small scales,
and for the first time we quantify the convergence of two independent isotropic particle loads (i.e. glass and CCVT) on halo properties.

\end{abstract}


\section{Introduction}\label{sec_intro}
Cosmological $N$-body simulations are one of the most crucial tools in cosmology to study the formation of structures in the universe \citep*[see e.g.][for reviews]{Frenk2012,Kuhlen2012}. To prepare an initial condition for a large-volume cosmological $N$-body simulation, the first step is to generate a homogeneous distribution of $N$ particles. This uniform particle distribution is usually named a pre-initial condition or a particle load\footnote{Note that we use these two terms interchangeably in this article.}.

There are several known methods to prepare a pre-initial condition: (i) The grid (or simple cubic lattice) load \citep[e.g.][]{Efstathiou1985} simply places $N$ particles in a regular grid, and it is uniform but anisotropic \citep*[see][for considerations on other related Bravais lattice configurations, e.g. body-centred cubic and face-centred cubic]{Joyce2009}. (ii) The glass method \citep{White1996} starts from a random distribution of $N$ particles, and then evolves the whole configuration under anti-gravity until it reaches an equilibrium state \citep*[see][for a similar consideration of relaxing smoothed particle hydrodynamics (SPH) particles at constant entropy under the influence of gas pressure until all particles have roughly equal SPH densities]{Couchman1995}. The glass pre-initial condition is homogeneous and isotropic. (iii) The Q-set method \citep{Hansen2007} partitions the space recursively using the quaquaversal tiling and places a particle in each quaquaversal tile. Note that the total particle number of a Q-set configuration is restricted to $2 \times 8^{N_{\rm iter}}$, with $N_{\rm iter}$ being the number of iterations. A Q-set configuration with finite particle number is not essentially isotropic, but it approaches the condition of isotropy in a statistical sense with large $N_{\rm iter}$ \citep[see][for detailed discussions]{Hansen2007}. (iv) Recently, \citet{Liao2018} adopts the capacity constrained Voronoi tessellation (CCVT) from computer graphics to generate a particle configuration which satisfies two constraints, i.e. the volume of each particle's Voronoi cell is equal and each particle resides in the centre-of-mass position of its Voronoi cell. It is shown in \citet{Liao2018} that a CCVT configuration is uniform and isotropic. The CCVT pre-initial conditions have been used in simulations of interacting dark energy models \citep[][]{Zhang2018}.

Quantifying and understanding the impacts of numerical setups on physical results is an essential task for simulation studies. Many previous studies have investigated the impacts of pre-initial conditions on the formation and evolution of large-scale structures by performing cosmological $N$-body simulations starting from grid, glass, and Q-set loads. For example, \citet*{Baugh1995} use cosmological simulations starting from grid and glass loads to study the discreteness effects on two-point correlation functions and power spectra for the matter distribution. These authors find that at early times, glass simulations show higher correlation functions on small scales than grid simulations, but such small discrepancies  become negligible at later times. These observations are qualitatively confirmed by \citet{Joyce2009}, who further quantify that the impacts of pre-initial conditions on matter power spectra at scales around the Nyquist frequency are at per cent levels (see also \citealt{Smith2003} and \citealt*{LHuillier2014}). Apart from the aforementioned two-point statistics, halo mass functions and pairwise velocities have also been shown to converge at a level of several per cent among simulations starting from grid and glass particle loads \citep[see e.g.][]{LHuillier2014,Liao2018}. \citet*{Schmittfull2013} show that at $z \la 3$, the bispectra measured from grid and glass simulations also converge at per cent levels. Different particle loads would also produce spurious structures in filaments in warm/hot dark matter simulations, these effects have been addressed by e.g. \citet{Gotz2002,Gotz2003} and \citet{Wang2007}. Especially \citet{Wang2007} disfavor the use of Q-set loads by showing that the spurious low-mass structures produced in hot dark matter simulations with the Q-set load are more abundant and more complex than the cases with grid and glass loads. Recently \citet*{Masaki2020} study the impacts of pre-initial conditions
on the anisotropic separate universe simulations, by studying the tidal response function caused by the supersurvey modes.
They show that the grid load can produce artificial features which can be seen until $z\sim 9$ while the simulations of glass loads are more stable and accurate over the interested range of scales. But the impacts of pre-initial conditions become negligible at $z \leq 3$ in their simulations.

In this work, with the newly introduced uniform and isotropic CCVT particle load, we revisit the study of numerical convergence for pre-initial conditions. \citet{Liao2018} has shown that the {\it large-scale} statistics (e.g. power spectrum, two-point correlation function, pairwise velocity and halo mass function) converge at per cent level among simulations starting from grid, glass and CCVT loads. Here we extend the work of \citet{Liao2018} to study the numerical convergence of pre-initial conditions on {\it small-scale} halo properties. 
We would like to quantify how well CCVT simulations converge to the widely used grid and glass simulations at small scales.
Especially, CCVT and glass are two independent isotropic pre-initial conditions, for the first time we are able to quantify the convergence of isotropic pre-initial conditions, and to compare it with the convergence between grid and glass which has been studied before. As we will show below, when looking at halo properties, CCVT simulations converge slightly better to glass simulations compared to their convergence to grid simulations.

The goal of this work is to study the numerical convergence (i.e. the agreement between different simulations) instead of the physical convergence (i.e. the agreement between the simulation results and the absolute physical results which are independent of numerical setups and parameters). Although the latter is of ultimate interest, currently we still lack analytical tools to precisely study the physical convergence in the nonlinear structure formation process \citep[but see e.g.][for the introduction of analytical tools to study the discreteness effects on the evolution of power spectrum]{Joyce2005,Marcos2006}.

The structure of this paper is as follows. We describe the details of our simulations, halo identifications and computations of halo properties in Section~\ref{sec_methods}. In Section~\ref{sec_results}, we present the convergence results of halo properties from simulations with different pre-initial conditions, and the environmental dependence of convergence results. We conclude and summarise in Section~\ref{sec_con}. Not to make the results too tedious, we list some examples and discussions on the numerical convergence of halo masses, centres and other properties in Appendices \ref{ap:mismatched_halo} and \ref{ap:evolution}.

\section{Numerical methods} \label{sec_methods}

\subsection{Simulations}

We use the \textsc{Gadget-2} code \citep[]{springel2005} to perform three cosmological $N$-body simulations starting from three different pre-initial conditions, i.e. grid, glass and CCVT\footnote{Note that we do not consider the Q-set pre-initial condition here because its particle number is restricted to some specific numbers, it is not ideally isotropic, and it has been shown to perform worse than either a grid or a glass load \citep{Wang2007}.}. The glass load containing $64^3$ particles is prepared using the \textsc{Gadget-2} code, while the CCVT pre-initial condition containing $64^3$ particles is from \citet{Liao2018} with a capacity parameter of $20^3$. Our adopted cosmological parameters are $\Omega_{\rm m}=0.3, \Omega_\Lambda=0.7, \sigma_8=0.9, h=0.7$ and $n_{\rm s}=0.96$. The initial conditions are generated at redshift $z=127$ using the \textsc{N-Genic} code and the input matter power spectra are from \citet[]{eisenstein1998}. Each simulation contains $N_{\rm p}=512^3$ dark matter particles in a periodic cube with a side length of $L_{\rm box}=200$ $h^{-1}\mathrm{Mpc}$. The mass resolution of each simulation is $m_{\rm p} = 4.96\times 10^{9} \: h^{-1}{\rm M}_\odot$. The gravitational softening length is $7.81 \: h^{-1}\mathrm{kpc}$ (i.e. approximately $1/50$ of the mean inter-particle separation), which is roughly the optimal softening length for the haloes studied in this work according to \citet{Zhang2019}. There are 135 output snapshots (with a time interval of $\sim 0.1$ Gyr) in total for each simulation.

Note that we input the same random seed in \textsc{N-genic} to generate the initial conditions for all three simulations, and thus we can cross-match haloes in these simulations which produce identical large-scale structures. We have also varied the random seed to perform two additional sets of simulation, and made sure that all our conclusions do not depend on simulation realizations. In the following sections, we will only present the results of the first realization set.

Also note that both glass and CCVT particle loads are not unique and they can have different realizations. To study the impacts from different realizations of glass/CCVT loads, we have performed two additional simulations starting from a different glass realization (denoted as `glass-1') and a different CCVT realization (denoted as `CCVT-1'). The glass-1 (CCVT-1) simulation and the aforementioned glass (CCVT) simulation only differ in the input particle load and otherwise they use the same parameters and setup.

\subsection{Halo identification and cross matching}\label{sec:halo_iden}

\begin{table} 
\centering
\caption{Number of haloes in different simulations.}\label{table_halo_num}
  \begin{tabular} {@{}cccc@{}}
  \hline
  Halo sample & Glass & Grid & CCVT\\
 \hline
   ALL & 2460 & 2481 & 2474\\
   \\
   MATCHED & 1192 & 1192 & 1192\\
   MISMATCHED & 1268 & 1289 & 1282\\
\hline
\end{tabular}
\end{table}

\begin{figure*} 
\centering\includegraphics[width=350pt]{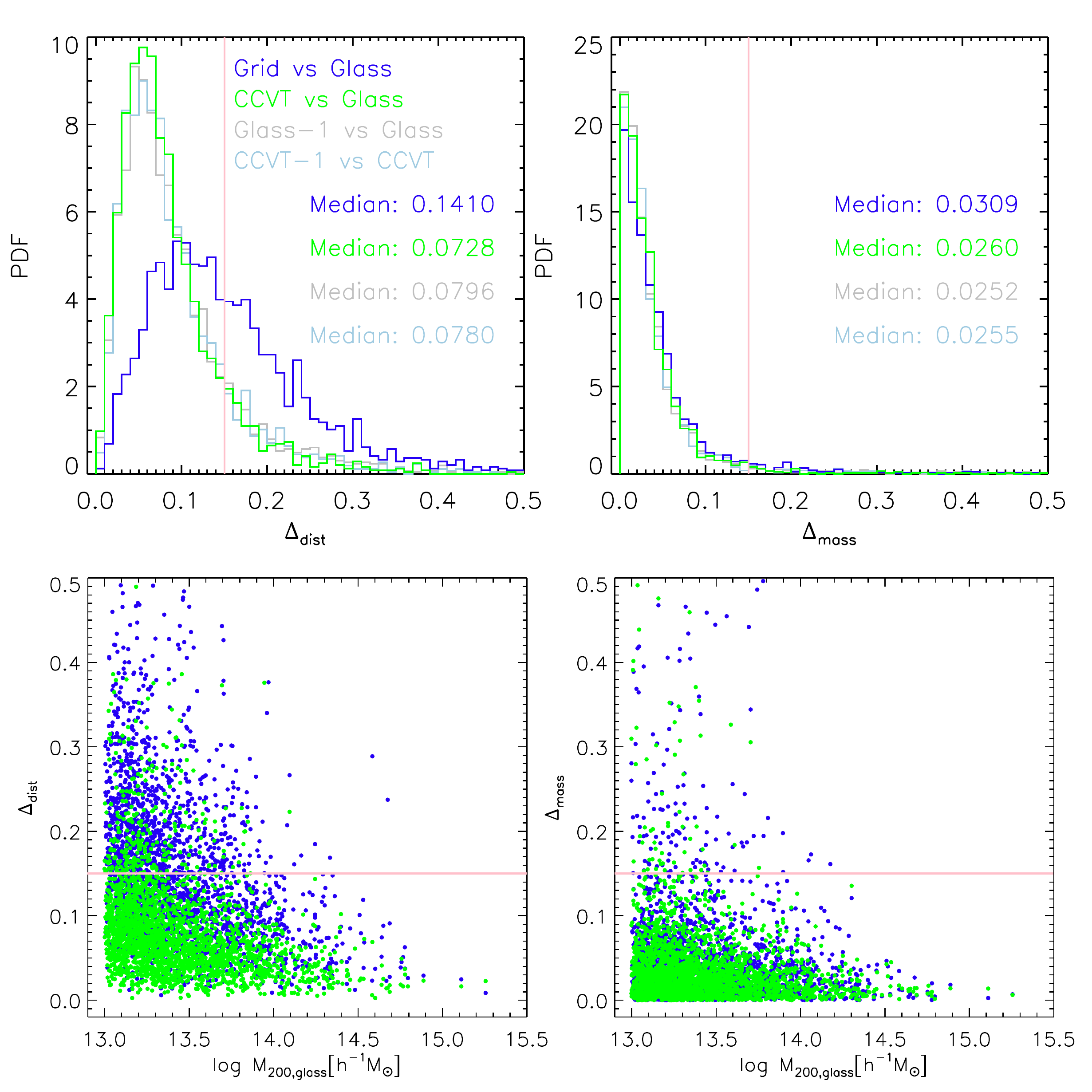}
\caption{Top: Probability distribution functions of $\Delta_{\rm dist}$ (left) and $\Delta_{\rm mass}$ (right). We use blue and green colours to distinguish results from grid and CCVT simulations. The median values of $\Delta_{\rm dist}$ or $\Delta_{\rm mass}$ are shown in each panel accordingly. The pink vertical lines mark the threshold of $0.15$ which we used to define MATCHED haloes. As a comparison, we use grey (light blue) lines to overplot the PDFs of the relative centre offset and relative mass difference for simulations starting from two different glass (CCVT) realizations. Bottom: Halo mass dependence of $\Delta_{\rm dist}$ (left) and $\Delta_{\rm mass}$ (right). Again, the grid and CCVT simulations are plotted with blue and green colours respectively. The pink horizontal lines mark the threshold of $0.15$.}
\label{fig:delta_pdf}
\end{figure*}

Dark matter haloes are identified by the friends-of-friends (FOF) algorithm \citep[]{Davis1985} with a linking length parameter of 0.2 times the mean inter-particle separation. We further use the \textsc{HBT+} halo finder code \citep[]{Han2018}, which is an improved version of the \textsc{HBT} code \citep[]{Han2012}, to identify subhaloes with at least 20 particles from the FOF groups and construct the subhalo merging history.

The halo centre is defined as the position of the most bound particle, $\bm{r}_{\rm min, pot}$, in each FOF group. The boundary of a halo is defined by the virial radius, $R_{200}$, within which the mean density is 200 times the cosmic critical density. In this article, we use $M_{200}$ and $N_{200}$ to denote the halo virial mass and the halo total particle number inside $R_{200}$ respectively. To achieve better numerical resolution for the halo sample, we only use haloes with $N_{200} \geq 2000$ (i.e. $M_{200} \geq 9.9 \times 10^{12} \: h^{-1}{\rm M}_\odot$) for analysis. In total we have 2460, 2481 and 2474 haloes in the glass, grid and CCVT simulations
respectively (i.e. the `ALL' sample in Table~\ref{table_halo_num}).

To have one-to-one comparisons between haloes from different simulations, we have tried to cross-match the $z=0$ haloes from three simulations by comparing their centre positions and virial masses. Taking glass haloes as reference, we first match haloes between the glass and grid simulations. Specifically, for each glass halo, we compute its relative centre offset and relative mass difference with respect to every grid halo,
\begin{equation}
    \Delta_{{\rm dist}, i} = \frac{\left|\bm{r}_{\rm min,pot}^{{\rm grid}, i} - \bm{r}_{\rm min,pot}^{\rm glass}\right|}{R_{200}^{\rm glass}},
\end{equation}
and
\begin{equation}
    \Delta_{{\rm mass}, i} = \frac{\left|M_{200}^{{\rm grid}, i} - M_{200}^{\rm glass}\right|}{M_{200}^{\rm glass}},
\end{equation}
where $\bm{r}_{\rm min,pot}^{{\rm grid}, i}$ and $M_{200}^{{\rm grid}, i}$ represent the centre and mass of the $i$th grid halo respectively, and $\bm{r}_{\rm min,pot}^{\rm glass}$ and $M_{200}^{\rm glass}$ denote the centre and mass of the targeted glass halo respectively. The matched grid halo of this glass halo is defined as the one with the minimum $\Delta_{{\rm dist}, i}^2 + \Delta_{{\rm mass}, i}^2$. Similarly, we can also match glass haloes to the CCVT haloes. We have also cross-matched the haloes in the simulations starting from different particle load realizations (i.e. glass-1 versus glass and CCVT-1 versus CCVT).

The probability distribution functions (PDFs) of $\Delta_{\rm dist}$ and $\Delta_{\rm mass}$ for all matched haloes in both grid and CCVT simulations are shown in the upper panels of Fig. \ref{fig:delta_pdf}. As a comparison, we also overplot the results from the haloes matched in simulations starting from different load realizations (i.e. grey lines for glass-1 versus glass, and light blue lines for CCVT-1 versus CCVT). Grid and CCVT haloes have fairly similar PDFs for $\Delta_{\rm mass}$, with a median of $\sim 0.03$, and this convergence in halo mass among different particle loads is fairly similar to that among different realizations of the same load class. However, for $\Delta_{\rm dist}$, CCVT haloes are more skewed to lower values comparing to grid haloes. The median values of $\Delta_{\rm dist}$ are 0.07 and 0.14 for CCVT and grid haloes, respectively. This indicates that CCVT haloes converge better to glass haloes in halo centres. Especially, the convergence in halo centres between CCVT and glass simulations is almost indistinguishable from that between different realizations of the glass (or CCVT) class. This implies that CCVT and glass loads share fairly similar properties in uniformity and isotropy.

If a glass halo has both its grid and CCVT counterparts satisfying
\begin{equation}
    \Delta_{\rm dist} \leq 0.15 \: {\rm and} \: \Delta_{\rm mass} \leq 0.15 
\end{equation}
at the same time, which indicates that the corresponding haloes in three simulations converge better than a level of $15\%$ in both halo centre positions and virial masses, we classify this glass halo and its matched grid/CCVT counterparts into the MATCHED halo sample. In total, there are 1192 MATCHED haloes in each simulation. Then, the remaining haloes in each simulation are referred as MISMATCHED haloes. See Table~\ref{table_halo_num} for the numbers of MATCHED and MISMATCHED haloes in different simulations. Note that we have followed similar criteria to define the matched and mismatched halo samples between glass-1 and glass (also CCVT-1 and CCVT) simulations. We will compare the convergence among different particle loads with the convergence between different realizations of the same load class from time to time in the following sections.

In the bottom panels of Fig. \ref{fig:delta_pdf}, we plot the mass dependence of $\Delta_{\rm dist}$ and $\Delta_{\rm mass}$. Most of MISMATCHED haloes are low-mass haloes. However, we can also find that even some massive haloes are poorly converged in simulations with different pre-initial conditions. For example, two grid haloes with masses $M_{200} > 10^{14.5}$ $h^{-1}{\rm M}_\odot$ have $\Delta_{\rm dist} > 0.15$ and thus they are classified as MISMATCHED haloes. We have examined the growth histories of these massive haloes, and found that they are usually either merging haloes which have their major merger events happened in significantly different paces in different simulations or haloes in the relaxing process after recent major merger events which are out-of-sync in different simulations. During these out-of-sync merging processes, the scaled centre offset and mass difference can be larger than the threshold of $15\%$ temporarily (see Appendix \ref{ap:mismatched_halo} for further details and discussions).

Note that, apart from the threshold of 0.15 which classifies about half of the haloes in ALL sample as MATCHED haloes, we have also used other thresholds to define the MATCHED halo sample (e.g. 0.1 and 0.2) and confirmed that our main conclusions in this paper are not affected by these choices. In the rest of this paper, we will only present the results using the threshold of 0.15.

\subsection{Computations of halo properties}

In this work, we will examine the impacts of pre-initial conditions on various halo properties at $z=0$ including halo concentration, spin, maximum rotation velocity, radius corresponding to the maximum rotation velocity, shape, one-dimensional velocity dispersion, formation time, and subhalo mass fraction. These physical quantities have been widely used in the literature to quantify halo structures and properties, and they are important for us to understand the formation of haloes. We summarise the computation procedures of these halo properties as follows.

(1) Halo concentration $c_{\rm NFW}$. For each halo, we compute its density profile in 20 equally logarithmic radius bins in the range of $[0.05 R_{200}, R_{200}]$, and fit the density profile with the Navarro-Frenk-White (NFW) model \citep[]{Navarro1996,Navarro1997},
\begin{equation}
    \rho (r) = \frac{\rho_{\rm s}}{(r/r_{\rm s})(1+r/r_{\rm s})^2},
\end{equation}
where $\rho_{\rm s}$ and $r_{\rm s}$ are free parameters. The halo concentration is then computed as $c_{\rm NFW} = R_{200} / r_{\rm s}$.

(2) Halo spin $\lambda$. Following \citet{Bullock2001}, the spin parameter is computed as
\begin{equation}
    \lambda = \frac{J}{\sqrt{2}M_{200} V_{200} R_{200}},
\end{equation}
where $J$ is the magnitude of the total angular momentum $\bm{J} = \sum_{i=1}^{N_{200}} m_{\rm p} (\bm{r}_{i} - \bm{r}_{\rm halo, COM}) \times (\bm{v}_{i} - \bm{v}_{\rm halo, COM})$, $\bm{r}_{i}$ and $\bm{v}_{i}$ are the $i$th particle's position and velocity vectors respectively, $\bm{r}_{\rm halo, COM}$ and $\bm{v}_{\rm halo, COM}$ denote the halo centre-of-mass position and velocity respectively, $V_{200} \equiv \sqrt{G M_{200} / R_{200}}$ is the halo rotation velocity at $R_{200}$ and $G$ is the gravitational constant.

(3) Halo maximum rotation velocity $V_{\rm max}$. $V_{\rm max}$ is the maximum of the halo rotation curve,
\begin{equation}
    V(r) = \sqrt{\frac{G M(<r)}{r}},
\end{equation}
where $M(<r)$ is the enclosed mass within the halo radius $r$. 

(4) $r_{\rm max}$. It marks the halo radius where the halo rotation curve reaches the maximum velocity (i.e. the aforementioned $V_{\rm max}$).

(5) Halo shape. To measure halo shape, we first compute the tensor
\begin{equation}
    I_{\alpha \beta} = \sum_{i=1}^{N_{200}} \left(r_{i,\alpha} - r_{\mathrm{min,pot}}^\alpha\right)\left(r_{i,\beta} - r_{\mathrm{min,pot}}^\beta\right),
\end{equation}
where $\alpha$ and $\beta$ refer to $x$, $y$ or $z$, $r_{i, \alpha}$ is the $\alpha$-component of the $i$th particle's position vector, and $r_{\rm min,pot}^\alpha$ is the $\alpha$-component of the halo centre position vector. We then compute the eigenvalues of $I_{\alpha\beta}$, $\lambda_1 \geq \lambda_2 \geq \lambda_3$. The lengths of the halo major, mediate and minor axes are given by $a=\sqrt{\lambda_{1}}$, $b=\sqrt{\lambda_{2}}$ and $c=\sqrt{\lambda_{3}}$, respectively. In this work, we mainly present the results of the minor-to-major axial ratio $c/a$. 

(6) Halo one-dimensional velocity dispersion $\sigma_v$. Following \citet{Evrard2008}, we compute the one-dimensional velocity dispersion as
\begin{equation}
    \sigma_v = \sqrt{\frac{1}{3N_{200}}\sum_{i=1}^{N_{200}}|\bm{v}_{i} - \bm{v}_{\rm halo, COM}|^2}.
\end{equation}

(7) Halo formation time $z_{\rm f}$. The halo formation time is defined as the redshift when the halo main progenitor first reaches half of the virial mass at $z=0$, i.e. $M_{200}(z = z_{\rm f}) = M_{200}(z=0) / 2$. In practice, we follow the main branch of a halo's merger tree, and find out the snapshot at which the mass of the main progenitor first exceeds $M_{200}(z=0) / 2$, then the formation time is computed with a logarithmic interpolation between this snapshot and the previous one. 

(8) Subhalo mass fraction $f_{\rm sub}$. For each halo, we add up the masses of all resolved subhaloes (i.e. with at least 20 particles) within the halo virial radius, $\sum M_{\rm sub}$, and compute the subhalo mass fraction as $f_{\rm sub} = \sum M_{\rm sub} / M_{200}$.

\section{Convergence results}\label{sec_results}

\subsection{Convergence of median halo properties}

\begin{figure*}
\centering\includegraphics[width=380pt]{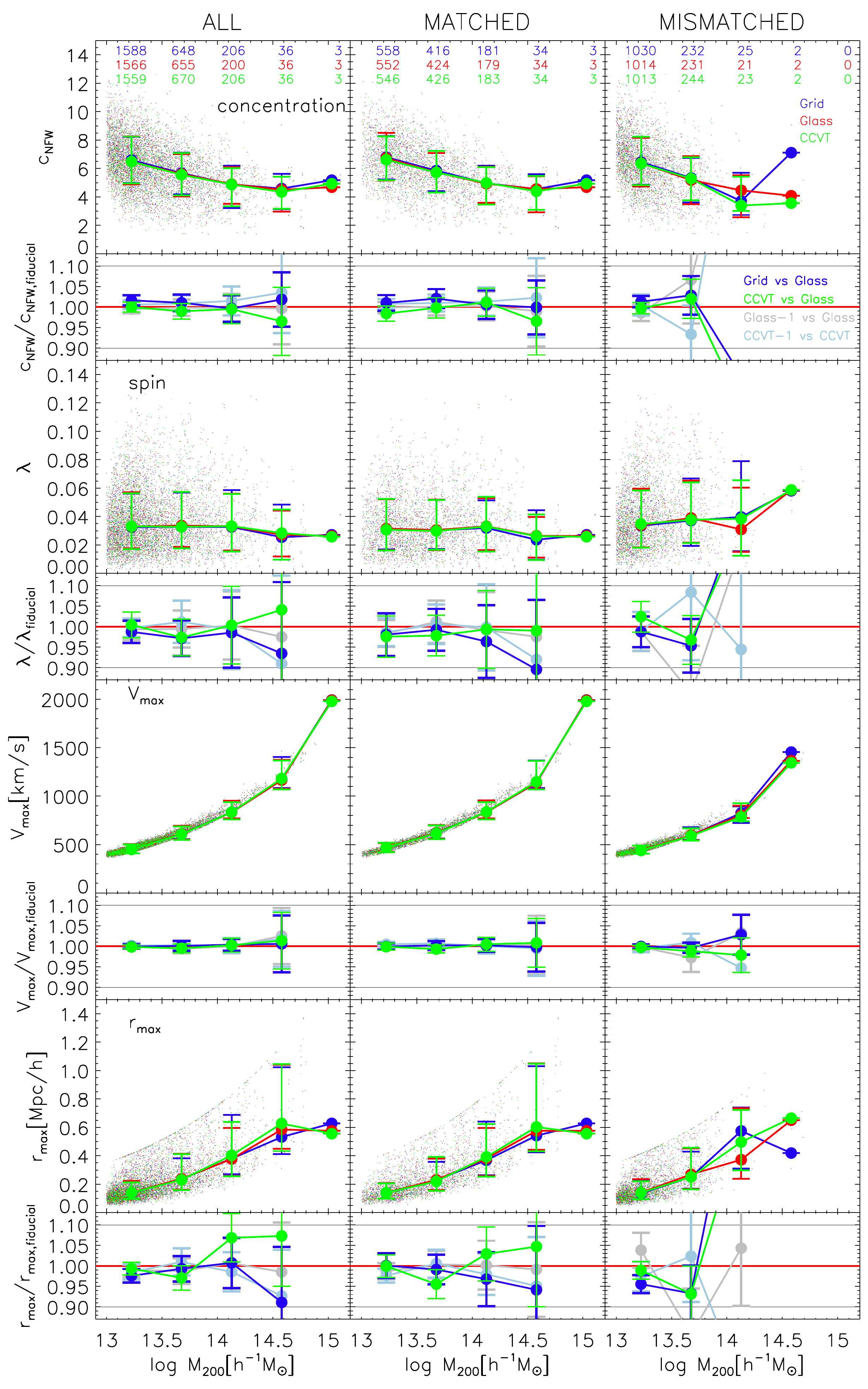}
\caption{Halo properties as a function of halo mass at $z=0$ for simulations with different initial loads. The grid, glass and CCVT simulations are plotted with blue, red and green colours. From top to bottom rows, we show halo concentrations, spins, $V_{\rm max}$ and $r_{\rm max}$ and their ratios of median values with respect to the glass simulation. From left to right columns, we show the results for haloes from ALL, MATCHED and MISMATCHED samples. In each panel, the scatter points mark all haloes from the sample, while the dots with error bars show the median values and the 16th and 84th percentiles in each mass bin. The grey horizontal lines in each ratio panel mark the 10\% convergence level. The errors of ratios are computed from the errors of median according to Eq. (\ref{eq:error_bar}). The numbers at the top of each column indicate the number of haloes within each mass bin. For comparison, in the ratio panels, we overplot the ratios of different median halo properties from glass-1 versus glass (CCVT-1 versus CCVT) simulations with grey (light blue) color.}
\label{fig1a}
\end{figure*}

\begin{figure*} 
\centering\includegraphics[width=380pt]{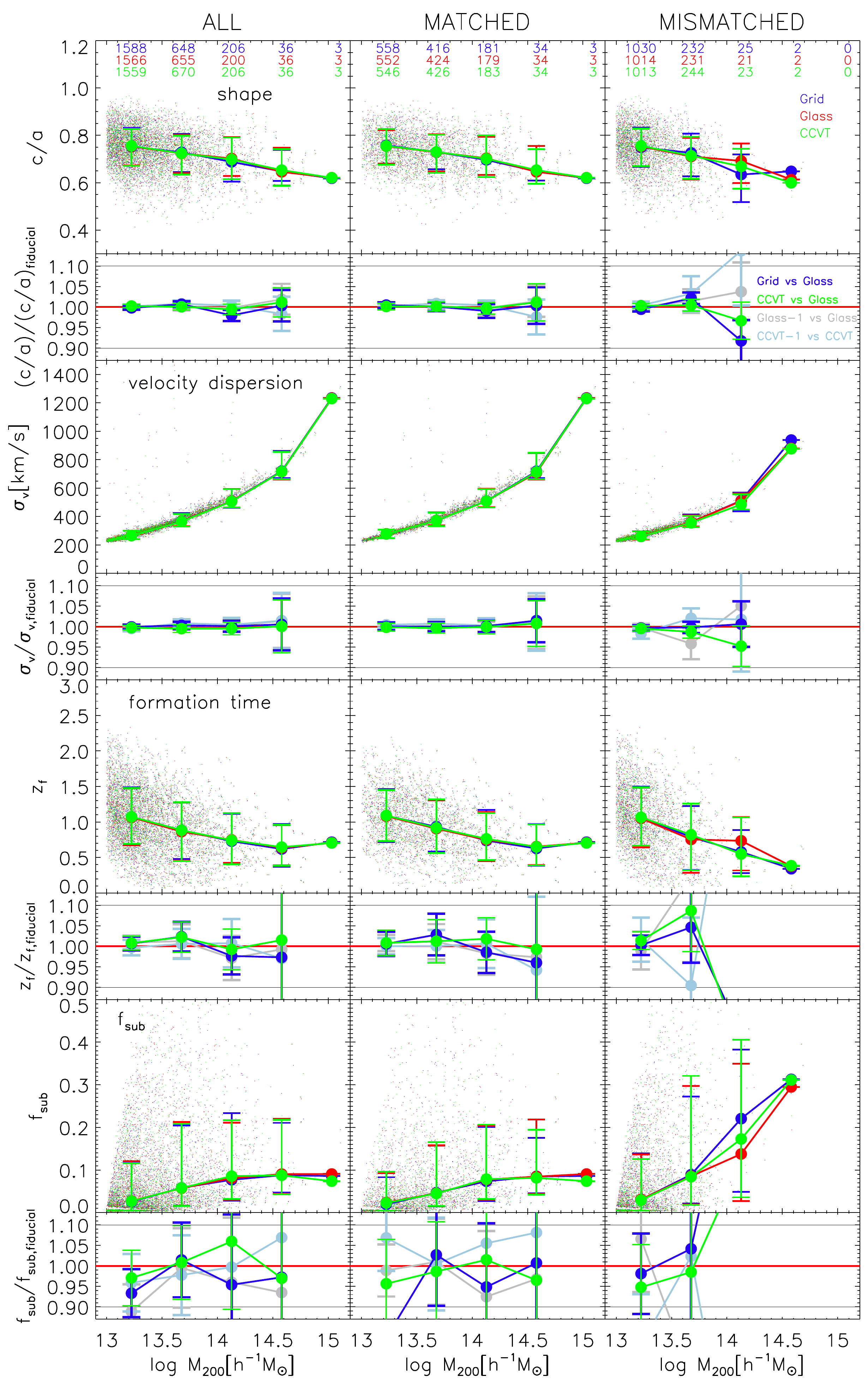}
\caption{Continued to Fig. \ref{fig1a}, for halo shape, velocity dispersion, formation time and subhalo mass fraction at $z=0$.}
\label{fig1b}
\end{figure*}

In this subsection, we study the convergence of median halo properties from simulations starting with different initial loads. In Figs~\ref{fig1a} and \ref{fig1b}, we plot the halo properties as a function of halo mass. From top to bottom, we show the halo concentration, spin, $V_{\rm max}$, $r_{\rm max}$, axial ratio $c/a$, velocity dispersion, formation time, and subhalo mass fraction. From left to right columns, we show the results from ALL, MATCHED and MISMATCHED halo samples. For each halo property, in the upper panel, the scatter points show the halo property - halo mass relation for all haloes in the sample while the large dots with error bars plot the median relation and its scatters in different mass bins; in the lower panel, we plot the ratio of medians with respect to the glass simulation. Note that, to have a robust estimation of the ratio of medians, we only show the ratios for the mass bins which have more than 20 haloes. The halo numbers in each mass bin are outlined at the top of the figures. We use blue, red and green colours to distinguish the results from the grid, glass and CCVT simulations respectively. To compare with the convergence between different realizations of the same load class, we overplot the ratio of medians from the glass-1 versus glass simulations and from the CCVT-1 versus CCVT simulations with grey and light blue colours, respectively.

Let us first focus on the left column in Figs~\ref{fig1a} and \ref{fig1b} to investigate the impacts of pre-initial conditions on ALL halo sample. It is easy to find that the median relations of halo property -- halo mass converge fairly well among grid, glass and CCVT simulations. Specifically, from the plots for ratios of median relations, both grid and CCVT simulations converge to glass simulation 
at a level of a few per cent (i.e. $\la 10\%$).
The numerical convergences among pre-initial conditions are particularly good for $V_{\max}$, $c/a$ and $\sigma_v$, which have convergence levels of $\la 2\%$. Quantitatively, similar convergence results are observed between different realizations of the same load class. From the plots of ALL halo sample, we confirm that the results of median halo properties in many previous studies, such as halo concentration - mass relations \citep[see e.g.][]{Neto2007, Maccio2007, Dutton2014}, converge at a level of a few per cent among different initial particle loads as well as different realizations of the same particle load class, and we are free to use grid, glass or CCVT loads to perform simulations when studying the median halo properties.

In the figures, we have also plotted the errors of ratios for different properties $x$, 
\begin{equation} \label{eq:error_bar}
\delta (x/x_{\rm glass}) = \sqrt{(\delta x/x_{\rm glass})^2 + (x \delta x_{\rm glass}/x_{\rm glass}^2)^2},
\end{equation}
where $\delta x$ and $\delta x_{\rm glass}$ denote the errors of median which are estimated by bootstrap resampling. The fluctuations of $x/x_{\rm glass}$ in different mass bins are within the range specified by these error bars, we therefore conclude that in general, the convergences of median halo properties do not depend on halo mass for our halo sample.

We further separate the whole ALL halo sample into MATCHED and MISMATCHED sub-samples, and plot their results in the middle and right columns in Figs~\ref{fig1a} and \ref{fig1b}. We can easily find that, for all median halo properties, the convergences are better for the MATCHED halo sample. For example, for median halo concentrations, the convergences among grid, glass and CCVT are better than $\sim 5\%$ for MATCHED haloes, while the convergence level exceeds $10\%$ in the third mass bin for MISMATCHED haloes. This indicates that overall the haloes which converge better in centre positions and virial masses tend to converge better in other halo properties.

For the subhalo mass fractions shown in the bottom panels of Fig. \ref{fig1b}, we can see that the median $f_{\rm sub}$ in different mass bins for MISMATCHED haloes tend to be significantly larger than those of MATCHED haloes. This is due to the fact that MISMATCHED haloes are usually merging haloes or haloes which have experienced recent major mergers, as we discussed in Section \ref{sec:halo_iden} and Appendix \ref{ap:mismatched_halo}.

As the median halo properties are fairly well converged among different simulations, we move on to look at the convergence of properties for individual haloes. To this end, in the following subsections, we will focus on the MATCHED halo sample consisting of 1192 haloes that are well one-to-one matched across three simulations.

\begin{figure*} 
\centering\includegraphics[width=500pt]{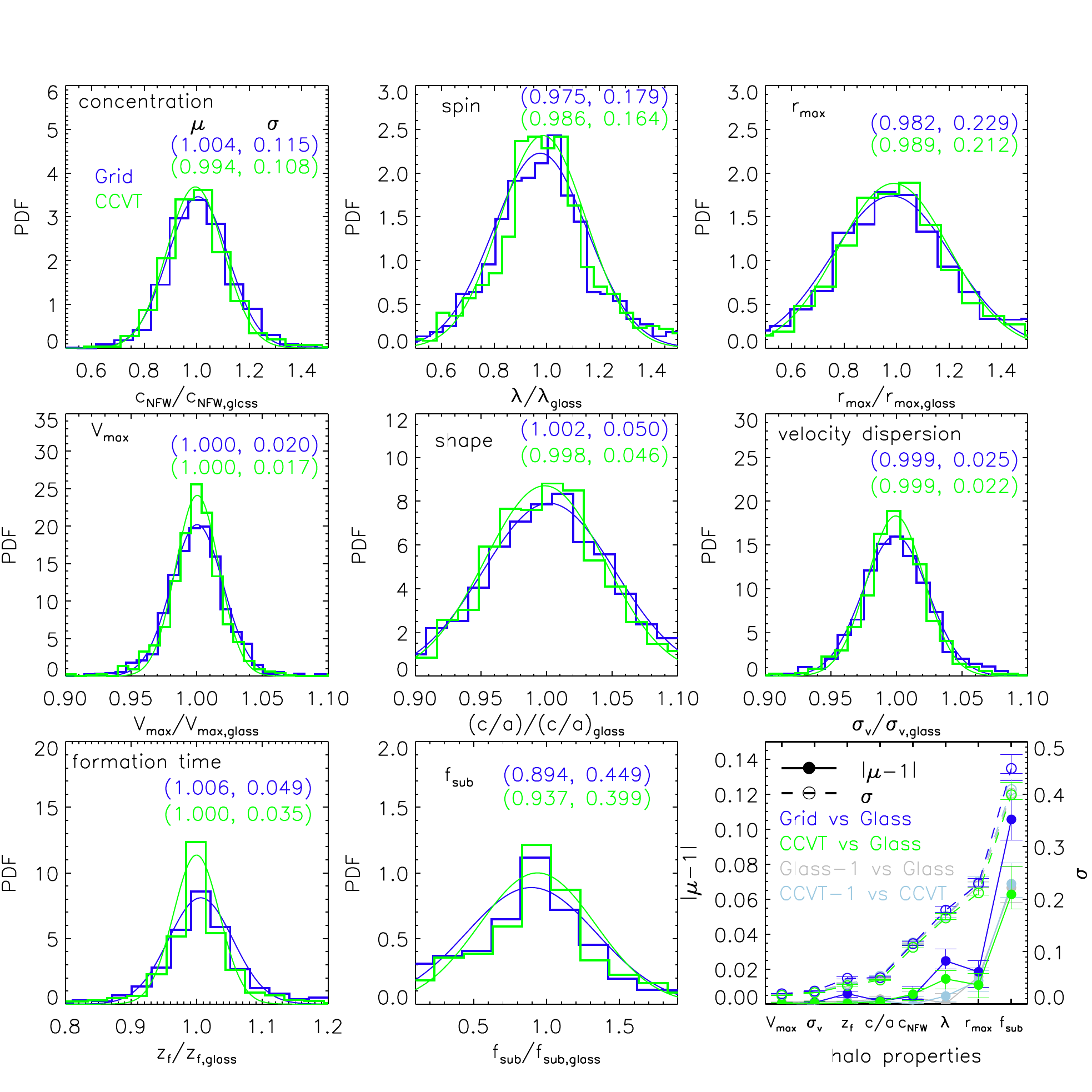}
\caption{Distributions of halo property ratios with respect to the glass simulation for MATCHED haloes at $z=0$. Grid and CCVT haloes are shown with blue and green colours respectively. Smooth curves are the best-fit Gaussian distribution functions, and the corresponding best-fit $\mu$ and $\sigma$ are given on the upper right corner of each panel. In the last panel, the best-fit $|\mu - 1|$ (solid, left $y$-axis) and $\sigma$ (dashed, right $y$-axis) are plotted versus different halo properties. To compare with the convergence between different realizations of the same load class, we overplot the results from glass-1 versus glass (CCVT-1 versus CCVT) simulations with grey (light blue) lines in the last panel. Note that the order of halo properties has been sorted according to the magnitude of $\sigma$ in the last panel. The error bars show the 16th and 84th percentiles estimated with the bootstrap resampling.}
\label{fig:matched_pdf_ccvt_grid}
\end{figure*}

\begin{figure*} 
\centering\includegraphics[width=400pt]{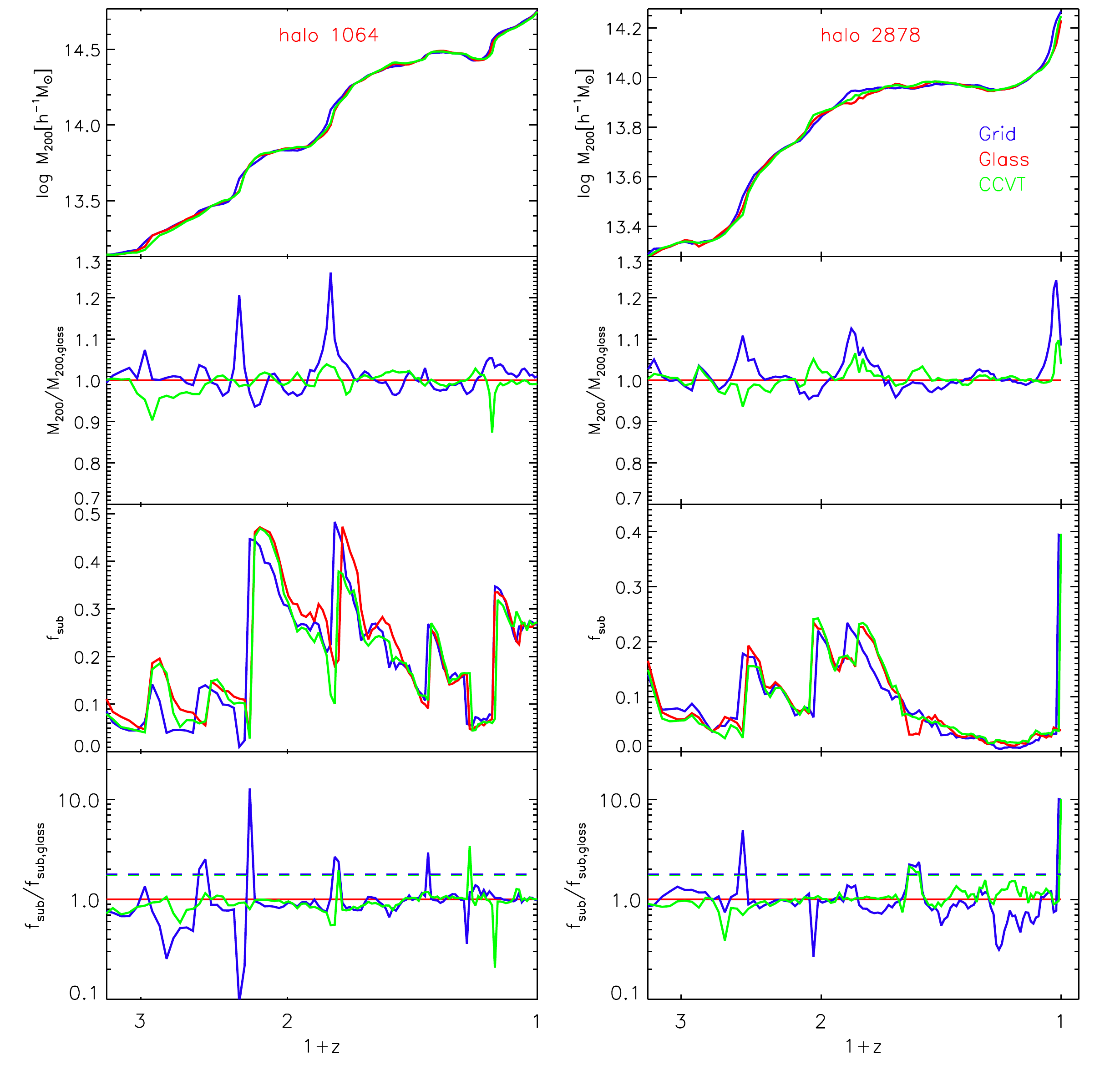}
\caption{Evolution of virial mass and subhalo mass fraction for Halo {\#}1064 (left) and Halo {\#}2878 (right) from MATCHED sub-sample. From top to bottom panels, we plot the time-evolution of virial masses, ratios of virial masses with respect to the glass halo, subhalo mass fraction, and ratios of subhalo mass fraction with respect to the glass halo. The grid, glass and CCVT simulations are plotted with blue, red and green lines respectively. The dashed lines in the bottom panels mark $f_{\rm sub}/f_{\rm sub, glass} = \mu + 2\sigma$. Note that the halo IDs mentioned here are from the glass simulation.}
\label{fig:fsub_evolution}
\end{figure*}

\begin{figure*} 
\centering\includegraphics[width=500pt]{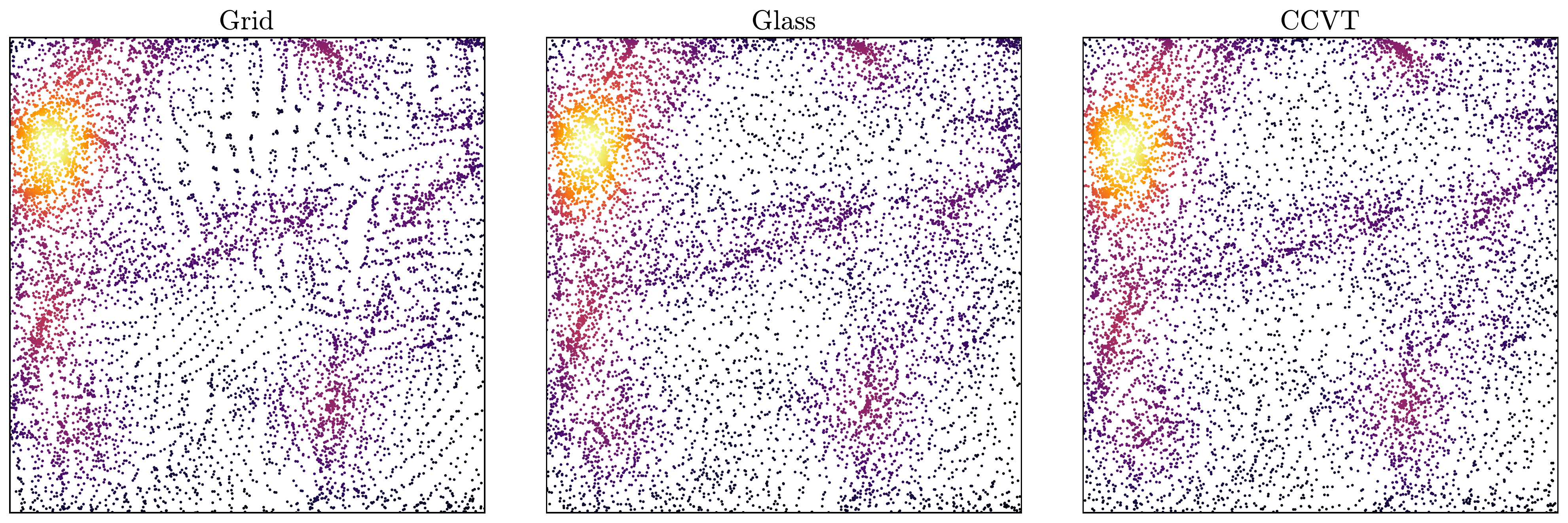}
\caption{Visualisation of $z=0$ void regions in simulations starting from grid, glass and CCVT particle loads (from left to right). In each panel, we project the particles within a $25~h^{-1} {\rm Mpc} \times 25~h^{-1} {\rm Mpc} \times 5~h^{-1} {\rm Mpc}$ slice onto the $xy$-plane. The colours denote the local density with black (yellow) colour representing low (high) density regions. In the grid simulation, we can clearly see some regular structures (i.e. particles like beads on a string) in the low density region.}
\label{fig:tendril_visua}
\end{figure*}

\subsection{Convergence on properties for MATCHED individual haloes}\label{subsec:individual}

In this subsection, we compare directly the halo properties of every matched halo pairs in the MATCHED halo sample. In Fig. \ref{fig:matched_pdf_ccvt_grid}, we compute the ratio of each property, $x/x_{\rm glass}$, for every grid-glass and CCVT-glass matched halo pair, and plot their PDFs. To better quantify their differences, we also fit each PDF with a Gaussian distribution,
\begin{equation}
    P(x/x_{\rm glass}) = \frac{1}{\sqrt{2\pi}\sigma} \exp{-\frac{(x/x_{\rm glass}-\mu)^2}{2\sigma^2}},
\end{equation}
with $\mu$ being the mean (or median) of the distribution and $\sigma$ being the dispersion. The best-fit $\mu$ and $\sigma$ are outlined in the upper-right corner of each panel. In the last panel, we summarise $|\mu - 1|$ and $\sigma$ for different halo properties.

We find that all PDFs can be well fitted with Gaussian distributions. All best-fit values of $\mu$ are close to one (i.e. $\la 10\%$), indicating that the median halo properties are well converged among different simulations. This is also the conclusion in the last subsection. We also note that the best-fit values of $\sigma$ are non-zero. Especially, the best-fit $\sigma$ of some PDFs, such as concentrations, spins, $r_{\rm max}$ and subhalo mass fractions, are larger than $10\%$. This indicates that although the median values converge at a level of a few per cent among different simulations, the relative differences for some individual haloes can be several tens of per cent or even larger.

To find out what happens for those extreme outliers whose properties converge poorly among different simulations, we look at MATCHED haloes with ratios deviating from $\mu$ larger than $2\sigma$. In Fig. \ref{fig:fsub_evolution}, we choose a typical example to illustrate why some haloes converge poorly in subhalo mass fractions which have the largest $\sigma$ among all properties discussed in this work. In the figure, we plot Halo {\#}1064 and Halo {\#}2878, which have similar masses ($\sim 10^{14.5}$ $h^{-1}{\rm M}_\odot$) at $z=0$, from the glass simulation and their matched counterparts in the grid and CCVT simulations. At $z=0$, the former halo has its subhalo mass fraction converged fairly well ($f_{\rm sub}/f_{\rm sub, glass} \approx 1$), while the subhalo mass fraction of the latter converges poorly ($f_{\rm sub}/f_{\rm sub, glass} \approx 10$). From top to bottom panels, we have plotted the time-evolution of their $M_{200}$, $M_{200}/M_{\rm 200, glass}$, $f_{\rm sub}$ and $f_{\rm sub}/f_{\rm sub, glass}$. We can clearly see from both haloes that if a halo experiences a major merger event, its subhalo mass fraction will increase significantly when the merging halo becomes a subhalo\footnote{The subhalo mass fraction then decreases gradually as the subhalo is tidally stripped by the host halo.}. However, the time when this merger event happens may be slightly different in simulations with different initial loads, and this leads both $M_{200}/M_{\rm 200, glass}$ and $f_{\rm sub}/f_{\rm sub, glass}$ deviate from 1 temporarily. Such behaviours can be observed almost in every major merger event in the formation history of both haloes. The reason for Halo {\#}2878 to have a very large $f_{\rm sub}/f_{\rm sub, glass}$ at $z=0$ is because it happens to experience a major merger event recently and its $f_{\rm sub}/f_{\rm sub, glass}$ is still in the temporarily deviating process. Note that at $z=0$, the ratios of $M_{200}/M_{\rm 200, glass}$ in both grid and CCVT simulations for this halo have already fallen below $1.15$ and thus it is classified as a MATCHED halo. Although we only illustrate one example here, similar behaviours can be seen in other haloes which have their subhalo mass fractions poorly converged at $z=0$.

For other halo properties, we also observe that when a merger event happens, and if this merger event happens in an out-of-sync manner in different simulations, the convergence can become poorer temporarily. Therefore, those haloes with their properties poorly converged at $z=0$ usually are experiencing merger events or experienced merger events recently and are still in the process of relaxation; see Appendix \ref{ap:evolution} for more details.

From the last panel of Fig. \ref{fig:matched_pdf_ccvt_grid}, we can also find that comparing to grid simulations, CCVT simulations converge slightly better to glass simulations, and the convergence between CCVT and glass simulations is almost indistinguishable from the convergence between different realizations of their own. In general, the best-fit $\mu$ for different halo properties from CCVT versus glass simulations are more close to 1, and the best-fit $\sigma$ from CCVT versus glass simulations are usually smaller than the ones from grid versus glass simulations. But we stress that their differences are tiny. This slightly better convergence between glass and CCVT is possibly due to the fact that both of them are isotropic while the grid configuration is anisotropic.

\begin{figure*} 
\centering\includegraphics[width=500pt]{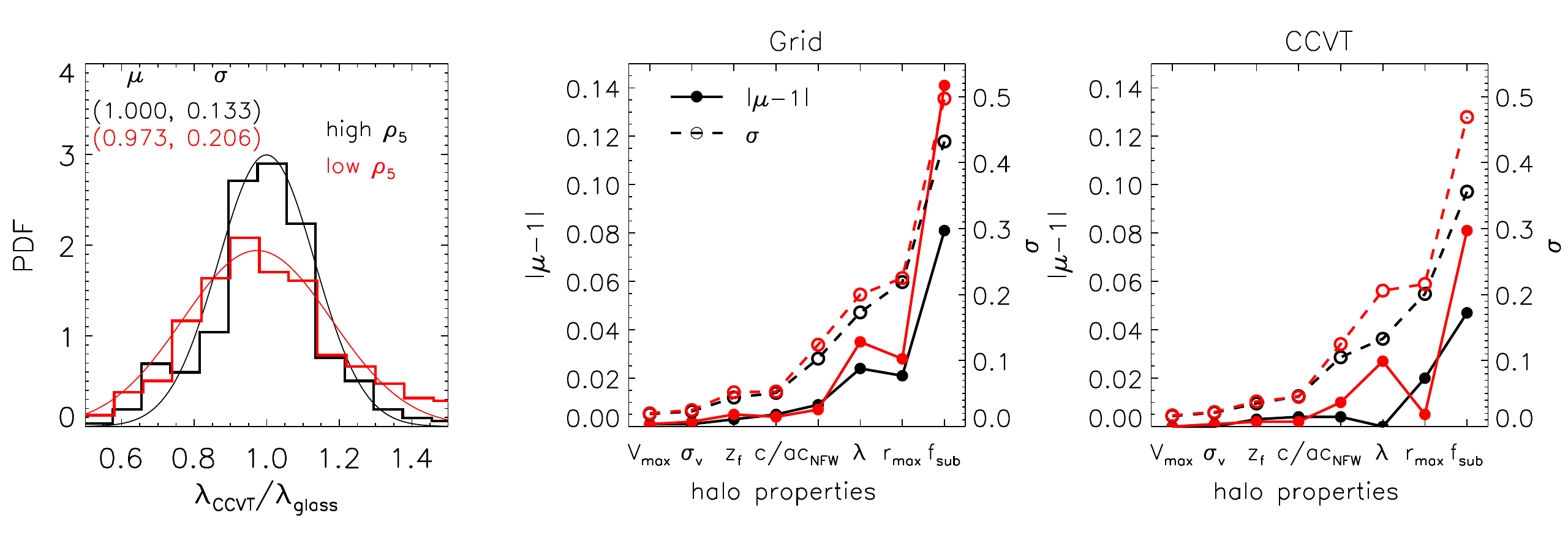}
\caption{Left: PDFs of $\lambda_{\rm CCVT}/\lambda_{\rm glass}$ for high-density (black) and low-density (red) haloes in the CCVT simulation. The PDFs are fitted with Gaussian distributions which are shown as smooth curves with corresponding colours. The best-fit $\mu$ and $\sigma$ of Gaussian distributions are outlined on the upper left corner. Middle: Best-fit $|\mu - 1|$ (solid, left $y$-axis) and $\sigma$ (dashed, right $y$-axis) for different halo properties in the grid simulation. High-density (Low-density) haloes are plotted with black (red) lines. Right: Similar to the middle panel, but for haloes in the CCVT simulation.}
\label{fig:environment_pdf}
\end{figure*}

\subsection{Role of environment}

It has been well known that the memory of pre-initial conditions is preserved in the void region of a simulation even at low redshifts \citep[e.g.][]{Baugh1995,White1996}. For example, in Fig. \ref{fig:tendril_visua}, we plot the $z=0$ particle distributions of void regions from simulations starting with different particle loads. In the left panel, we can clearly see some regular structures (i.e. particles like beads on a string) in low-density regions, which originate from the initial regular grid load. In contrast, we do not see such regular structures in glass and CCVT simulations whose initial particle loads are isotropic.

A natural and interesting question is: whether these large-scale environmental dependencies due to pre-initial conditions could affect the numerical convergence on halo properties? To answer this question, for both grid and CCVT simulations, we classify haloes from the MATCHED sample into high-density or low-density environments by computing their environmental number densities \citep{Dressler1980},
\begin{equation}
    \rho_n = \frac{3n}{4\pi D_n^3},
\end{equation}
where $D_n$ is the distance to the $n$-th nearest neighbour. In this work, we adopt $n=5$ and consider all isolated haloes containing more than 100 particles within their virial radii when searching for their 5th nearest neighbour. We sort the MATCHED haloes in each simulation according to their $\rho_5$, and classify the first one-third of haloes (397 haloes) with larger $\rho_5$ as high-density haloes and the last one-third of haloes (397 haloes) with lower $\rho_5$ as low-density haloes.

For both grid and CCVT simulations, we compute the PDFs of the ratios with respect to the glass simulation for different halo properties. We also fit the PDFs with Gaussian distributions to obtain the best-fit $\mu$ and $\sigma$. In the left panel of Fig. \ref{fig:environment_pdf}, we show an example of the PDFs of $\lambda_{\rm CCVT}/\lambda_{\rm glass}$ for both high-density haloes and low-density haloes in the CCVT simulation. We can clearly see that for halo spins, high-density haloes tend to have their $\mu$ more close to 1 and to have a smaller $\sigma$, indicating that high-density haloes have their spins converged better among simulations with different initial loads.

We summarise the environmental dependence of $|\mu - 1|$ and $\sigma$ for all halo properties in the middle (right) panel of Fig. \ref{fig:environment_pdf} for grid (CCVT) haloes. In general, haloes in high-density environments show slightly better convergence in properties (especially for concentrations, spins, $r_{\rm max}$, and subhalo mass fractions) among different simulations, i.e. haloes residing in high-density environments tend to have $\mu$ closer to 1 and to have smaller $\sigma$. This reflects that haloes in high-density environments tend to experience a higher degree of non-linear evolution and thus retain less memory of pre-initial conditions. Again, we stress that the environmental dependence for the convergence is only a weak effect. Note, we have checked that this environmental dependence of convergence can also be seen in the glass-1 versus glass and CCVT-1 versus CCVT simulations.

\section{Conclusions and discussions}\label{sec_con}

In this work, we study the numerical convergence of pre-initial conditions on dark matter halo properties, by using a set of cosmological $N$-body simulations starting from initial conditions with identical random phases but with different particle loads (i.e. grid, glass, and CCVT). Our conclusions can be summarised as follows:

(i) The median halo properties converge fairly well among simulations with different initial loads and among simulations with different realizations of the same glass/CCVT load class, i.e. 
at a convergence level of a few per cent.
The convergence is particularly good for $V_{\rm max}$, axial ratios and velocity dispersion which have convergence levels of $\la 2 \%$.

(ii) Haloes in the MATCHED sub-sample, which have their scaled centre offsets and mass differences $\leq 15\%$ across three simulations, generally converge better in their median properties than the remaining MISMATCHED haloes.

(iii) Although the median halo properties converge at levels of a few per cent among different simulations, the properties of some individual MATCHED haloes sometimes can be poorly converged (i.e. at convergence levels of several tens of per cent or even larger),
especially for halo concentrations, $r_{\rm max}$, spins and subhalo mass fractions. By tracing the growth histories of these poorly converged haloes, we find that they are either merging haloes or haloes have experienced recent merger events and still in the relaxation process, and the merging processes happen in an out-of-sync manner in different simulations. These out-of-sync merging events can lead to poor convergences temporarily.

(iv) Comparing with the grid simulation, the halo centre and other halo properties from the CCVT simulation converge slightly better to those from the glass simulation, and the convergence between CCVT and glass simulations is comparable to that between different realizations of the same glass/CCVT load class. This is possibly due to the fact that both glass and CCVT loads are isotropic while the grid load is anisotropic.

(v) Haloes in high-density environments tend to have their properties converge slightly better than haloes in low-density environments. This is possibly because haloes in high-density environments have a higher degree of non-linear evolution and retain less memory of initial loads.

Together with the convergence results of CCVT simulations on large-scale statistics from \citet{Liao2018}, 
by further investigating the numerical convergence of pre-initial conditions on small-scale halo properties, in this study we have confirmed that CCVT loads 
behave as well as the traditional grid and glass loads in cosmological $N$-body simulations.
Especially, for the first time, we are able to quantify the convergence between the two independent isotropic pre-initial conditions, i.e. CCVT and glass.

Our study also confirms that the results of median halo properties in many previous studies (e.g. halo concentration - mass relations, mass dependence of other halo properties, etc.) converge fairly well with different initial particle loads, and we are free to use grid, glass or CCVT loads to perform simulations when studying median halo properties.

 At the same time, we notice that for some individual haloes 
which are merging or have experienced merger events recently,
the relative differences of their halo properties among simulations with different particle loads can be temporarily several tens of per cent, even with the same random seed to generate the initial conditions. This cautions that we should check the numerical convergence carefully when we study individual haloes or a small sample of haloes in simulations, for example, in zoom-in simulations. Especially, if a zoomed individual halo experiences recent major merger events, its halo properties may not be well converged regarding the used pre-initial conditions. This is worth detailed and careful study in future works.

In this work, we only focus on the numerical convergence of pre-initial conditions, i.e. the agreement between simulations starting from different particle loads. The ultimate question on convergence is: how do simulation results converge to the absolute physical results? Although we still lack the precise analytical tools to answer this question, there have been some attempts in this direction. For example, \citet{Joyce2005} introduce the dynamical matrix formalism from solid state physics to study the discreteness effects from grid-like particle systems, and show that comparing to the theoretical fluid evolution, the growth of power spectrum for a grid particle system is suppressed at scales near the Nyquist frequency \citep[see further work in e.g.][]{Marcos2006,Joyce2007,Joyce2009}. Recently, \citet{Garrison2016} have developed a method to correct this effect on grid initial conditions. It will be interesting to apply this formalism to study the discreteness effects in glass and CCVT particle systems.

\section*{Acknowledgements}
We thank the anonymous referee for a very constructive report which helped improve this manuscript. We thank Liang Gao for useful discussions. We acknowledge support the from NSFC grants (Nos 11903043, 11988101). SL acknowledges the support by the European Research Council via ERC Consolidator Grant KETJU (no. 818930). JZ acknowledges support from IBS under the project code, IBS-R018-D1.

\section*{Data availability}
The simulation data underlying this article will be shared on reasonable request to the corresponding author. The CCVT particle load data used in this work is publicly available at \href{https://github.com/liaoshong/ccvt-preic-data}{https://github.com/liaoshong/ccvt-preic-data}.

\appendix

\section{Why do some massive haloes converge poorly in virial masses or centres?}\label{ap:mismatched_halo}

In this appendix, we study why some massive haloes converge poorly in halo masses or centres and why they are classified into the MISMATCHED sub-sample. As an example, we look at the most massive halo in the MISMATCHED sub-sample, which is numbered as Halo {\#}9812 and has a virial mass of $4.7 \times 10^{14}$ $h^{-1}{\rm M}_\odot$ in the glass simulation. We have matched its counterparts in the grid and CCVT simulations according to the procedures outlined in Section \ref{sec:halo_iden}. For both grid and CCVT simulations, their virial masses converge quite well to the glass simulation (i.e. $\Delta_{\rm mass} < 2.68\%$ and $< 1.42\%$ respectively). However, for the grid simulation, Halo {\#}9812 has $\Delta_{\rm dist} = 23.74 \%$, which makes it a MISMATCHED halo. In the CCVT simulation, $\Delta_{\rm dist}$ is $1.47 \%$.

\begin{figure} 
\centering
\includegraphics[width=240pt]{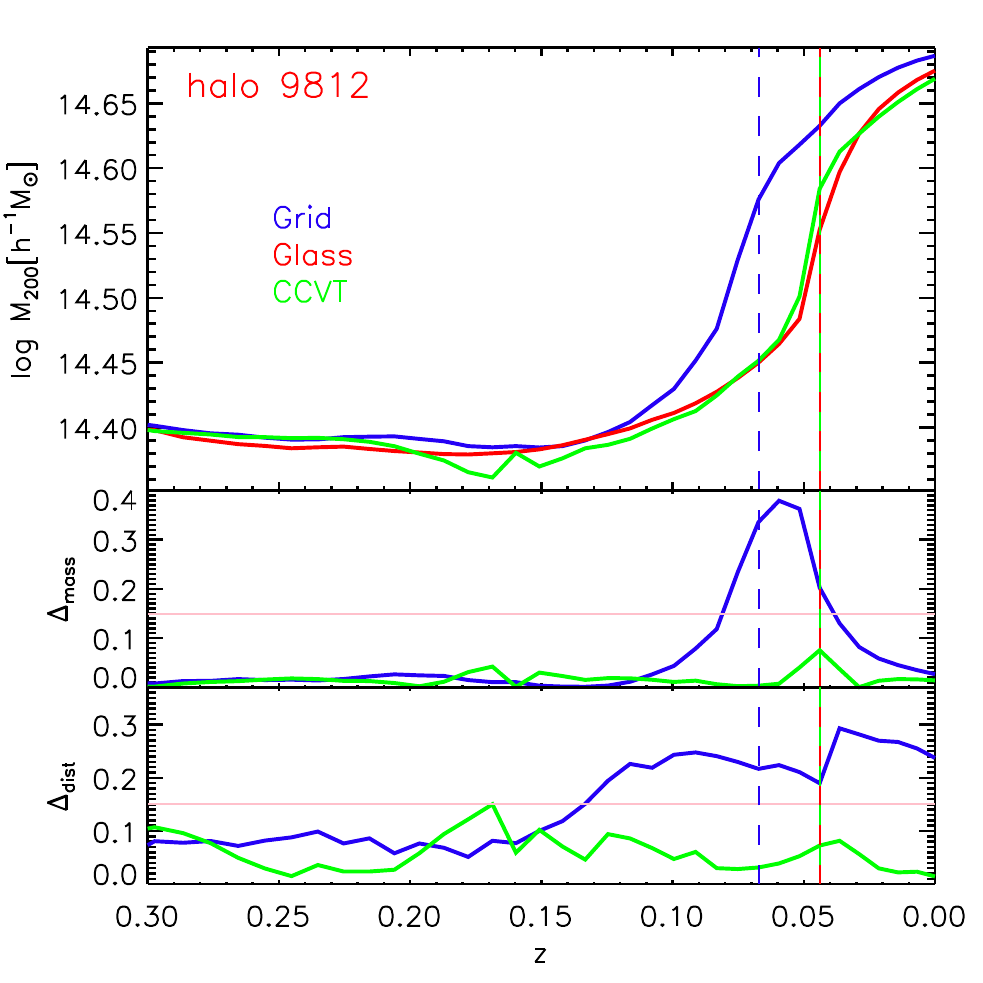}
\caption{The time evolution of properties for Halo {\#}9812 from the MISMATCHED sub-sample. From top to bottom panels, we plot its mass growth history, time evolution of $\Delta_{\rm mass}$ and $\Delta_{\rm dist}$. The haloes from grid, glass and CCVT simulations are shown with blue, red and green curves respectively. The vertical dashed lines mark the merging redshift, which is defined as the redshift of the snapshot in which the centre of the merging halo first moves into the virial radius of the main halo. In the lower two panels, the pink horizontal lines mark the threshold (15\%) that we use to define the MATCHED halo sample.}
\label{fig:mismatched_massive_halo}
\end{figure}

To understand why there is a large centre offset for Halo {\#}9812 between the grid and glass simulations, we have traced the evolution of the matched haloes along the main branch of their merger trees. In Fig. \ref{fig:mismatched_massive_halo}, we plot the mass growth history and the time-evolution of $\Delta_{\rm mass}$ and $\Delta_{\rm dist}$ from $z=0.3$ to $0$ for the matched haloes in three simulations.

\begin{figure*} 
\centering
\includegraphics[width=400pt]{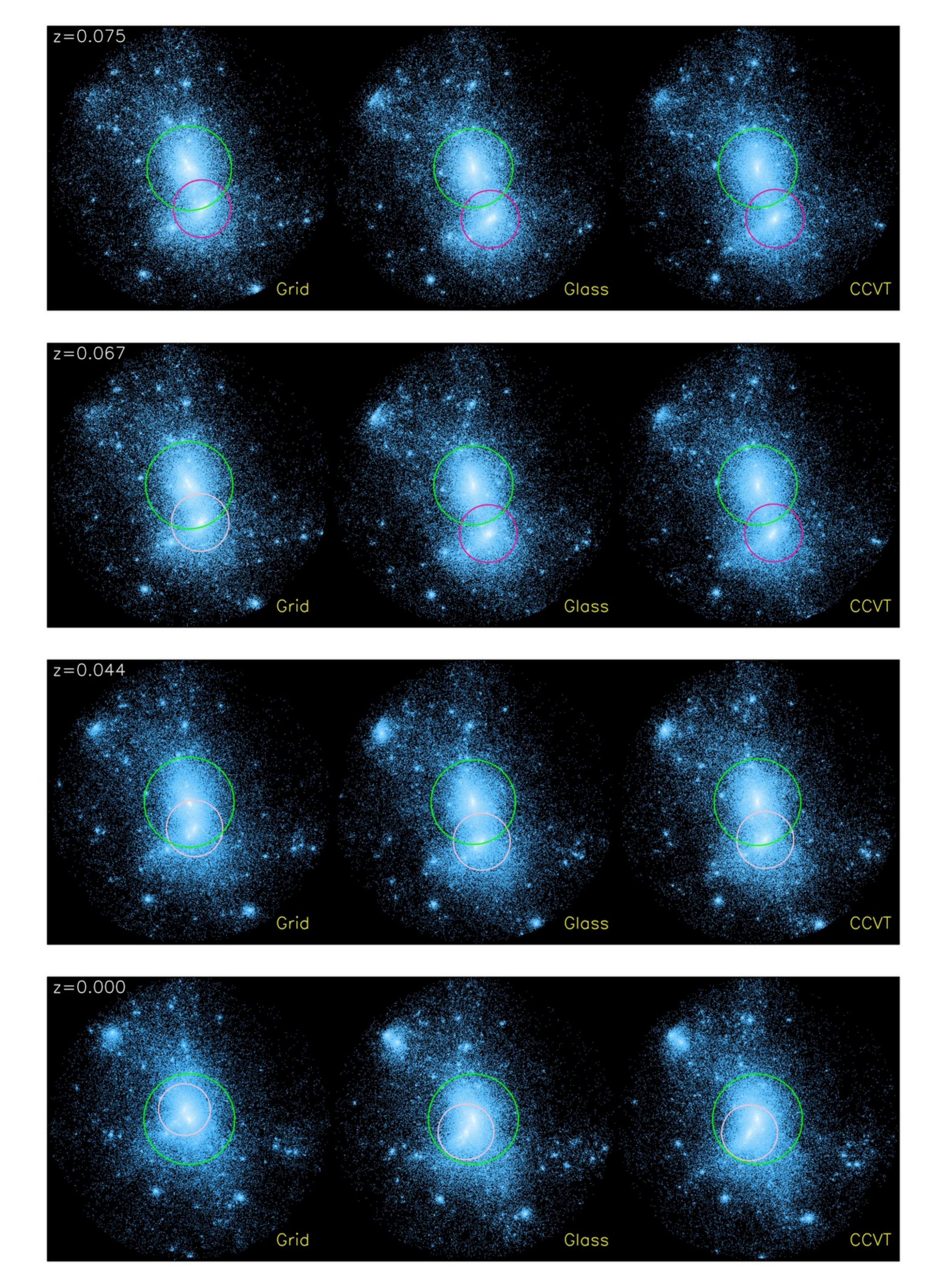}
\caption{Illustration of the merging process for Halo {\#}9812. From top to bottom rows, the projected density distributions centred on the halo main progenitor at $z=0.075, 0.067, 0.044$ and $0$ are shown. From left to right columns, we plot the visualisations from the grid, glass and CCVT simulations. The virial radius of the main progenitor is marked with a green circle centring the halo centre. The virial radius of the incoming halo is marked with a purple (pink) circle before (after) it becomes a subhalo. }
\label{fig:mismatched_evolution_halo_0}
\end{figure*}

From Fig. \ref{fig:mismatched_massive_halo}, it is evident that Halo {\#}9812 experiences a major merger event (with a merger mass ratio of $1:1.83$) recently at $z<0.1$. Prior to this merger event, both $\Delta_{\rm mass}$ and $\Delta_{\rm dist}$ are within the threshold of $15\%$ for both grid and CCVT simulations. In the grid simulation, this major merger event happens significantly ($\sim 0.3$ Gyr) earlier than those in the glass and CCVT simulations. To illustrate this process more clearly, we plot the evolution of the projected matter distribution centred on the main halo progenitor in Fig. \ref{fig:mismatched_evolution_halo_0}. At $z=0.067$, the centre of the incoming halo has already moved into the virial radius of the main halo in the grid simulation while the corresponding incoming halo is still outside the main halo in glass and CCVT simulations. In the later evolution, this subhalo in the grid simulation keeps moving at a different pace from that in the glass/CCVT simulation, affecting the potential well of the main halo differently. This different pace in the merging process leads both $\Delta_{\rm mass}$ and $\Delta_{\rm dist}$ in the grid simulation to larger deviations temporarily, as shown in Fig. \ref{fig:mismatched_massive_halo}. At $z=0$, as the incoming halo has already merged into the main halo in all three simulations, $\Delta_{\rm mass}$ drops back to a few per cent. However, as the halo is still relaxing, $\Delta_{\rm dist}$ in the grid simulation is still larger than $15\%$.

We have also looked at the evolution history of other massive haloes with $\Delta_{\rm mass} > 15\%$ or $\Delta_{\rm dist} > 15\%$, and found that they have similar evolutionary behaviours. Grid/CCVT haloes with $\Delta_{\rm mass} > 15\%$ are usually merging haloes which have their merging events happened significantly earlier or later than the glass counterparts. Grid/CCVT haloes with $\Delta_{\rm dist} > 15\%$ are usually haloes in the relaxing process after a recent major merger event which happened in significantly different pace from the glass counterparts.

Note that we have also compared the evolution history of some massive mismatched haloes between the glass-1 and glass (also CCVT-1 and CCVT) simulations, and we observed similar behaviours discussed above.

\section{Evolution of halo properties for outliers in the MATCHED sample}\label{ap:evolution}

\begin{figure*} 
\centering
\includegraphics[width=500pt]{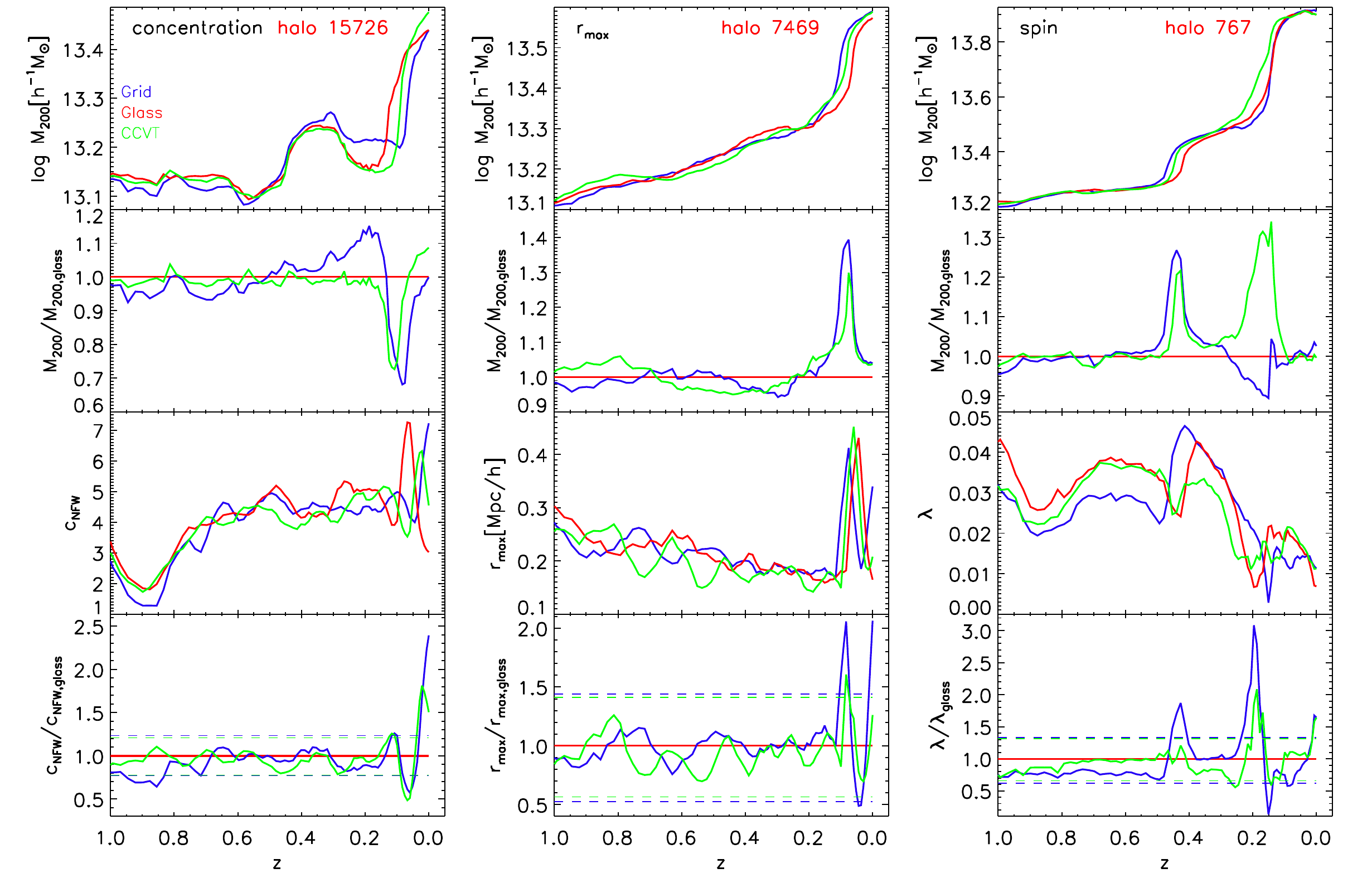}
\caption{Randomly selected examples from the MATCHED sub-sample, for the evolution of haloes whose $x/x_{\rm glass}$ at $z=0$ deviate from $\mu$ larger than $2\sigma$. From left to right columns, we show evolution examples for halo concentration, $r_{\rm max}$ and spin. In each column, from top to bottom panels, we plot the mass accretion history $M_{200}(z)$, the ratio of $M_{200}(z)/M_{\rm 200, glass}(z)$, property $x(z)$, and the property ratio $x(z)/x_{\rm glass}(z)$. Note that in this plot, the evolution of halo properties has been slightly smoothed (i.e. averaged over every five adjacent snapshots) in order to reduce noise. In the bottom panels of each column, the horizontal dashed lines mark the values of $\mu \pm 2\sigma$. Haloes from grid, glass and CCVT simulations are shown with blue, red and green colours respectively.}
\label{fig:property_example}
\end{figure*}

\begin{figure*} 
\centering
\includegraphics[width=500pt]{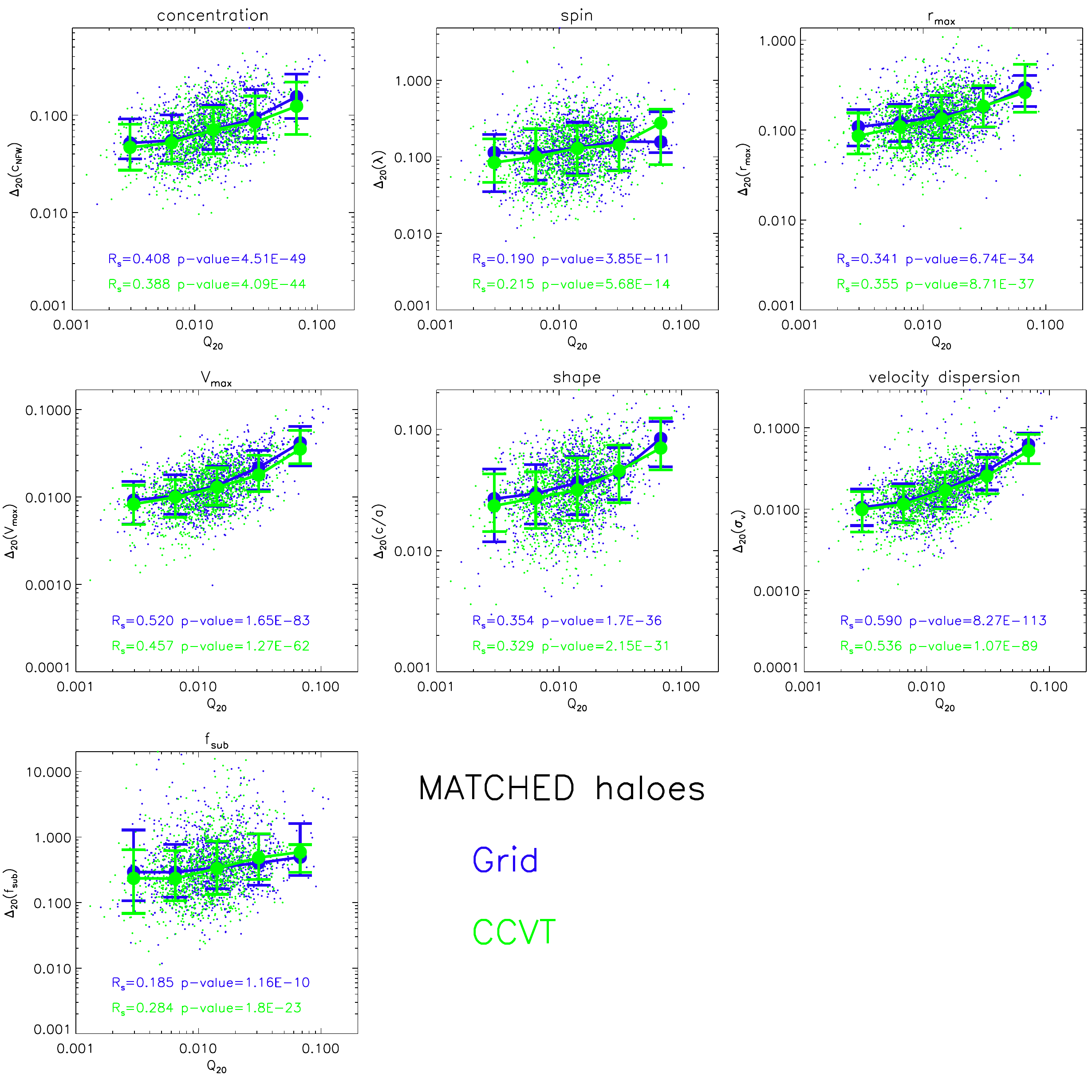}
\caption{Correlations between $Q_{20}$ and $\Delta_{20}(x)$ for different halo properties. From top to bottom and from left to right, we show the halo properties of concentration, spin, $r_{\rm max}$, $V_{\rm max}$, axial ratio, velocity dispersion and subhalo mass fraction. Grid haloes and CCVT haloes are marked with blue and green colours respectively. The scattering dots plot the single haloes, while the large dots with error bars show the median values and the 16th and 84th percentiles in five $Q_{20}$ bins. To have a robust estimation of the median values, we require that each $Q_{20}$ bin contains $\ga 20$ haloes. The Spearman's rank coefficients $R_{\rm s}$ and p-values are given in the lower part of each panel.}
\label{fig:Q20_D20_correlation}
\end{figure*}

In Section \ref{subsec:individual}, we have illustrated that during the processes of major mergers, if there is a large time lag among different simulations, the subhalo mass fraction can be poorly converged temporarily. In this appendix, we look at the evolution of other halo properties for some randomly selected haloes.

From Fig. \ref{fig:matched_pdf_ccvt_grid}, we know that in the MATCHED sample, apart from subhalo mass fractions, three other halo properties (i.e. concentrations, spins and $r_{\rm max}$) have $\sigma$ for their $x/x_{\rm glass}$ ratios larger than $10\%$. From each of these three halo properties, we look at the haloes whose ratios deviate from $\mu$ larger than $2\sigma$, randomly select a halo as an example, and plot the evolution of halo properties in Fig. \ref{fig:property_example}.

Both the $z=0$ concentrations of Halo {\#}15726 (left column of Fig. \ref{fig:property_example}) in grid and CCVT simulations are significantly different from that in glass simulations. This is associated with a major merger event happened after $z \sim 0.2$. Haloes in the glass simulation merged slightly earlier than the counterparts in the grid and CCVT simulations, which leads to temporary large mass deviations among different simulations. At the same time, the halo density profiles evolve rapidly and the fitted concentrations\footnote{Note that we have checked that the NFW profile still provide a relatively good description for the simulated density profile during the merging process for this halo.} oscillate with large amplitudes during the merging and relaxation processes. The out of sync concentration evolution leads to very poor convergence among three simulations after $z \sim 0.1$. Note that Halo {\#}15726 experienced another major merger event during the period from $z \sim 0.6$ to $z \sim 0.4$. However, as the merger happens 
almost synchronously in three simulations, the halo concentration converges fairly well during this merging process.

For Haloes {\#}7469 and {\#}767 shown in Fig. \ref{fig:property_example}, we can also observe similar evolution behaviors for their $r_{\rm max}$ and spins. If there are large differences for the mass growth history among different simulations, the convergences of $r_{\rm max}$ and spins can become poor temporarily. 

After illustrating examples of individual haloes, to look at the statistical correlation between the convergence in haloes' mass accretion history and the convergence in halo properties, we introduce two quantities, $Q_{N_{\rm s}}$ and $\Delta_{N_{\rm s}}(x)$, which are defined as follows.
\begin{equation}
    Q^2_{N_{\rm s}} \equiv \frac{1}{N_{\rm s}}\sum_{i=1}^{N_{\rm s}} \left[\log_{10}{M_{200}(z_i)} - \log_{10}{M_{\rm 200, glass}(z_i)}\right]^2
\end{equation}
measures the averaged difference of mass accretion histories for a halo from the grid (or CCVT) simulation and its counterpart from the glass simulation. Here, $N_{\rm s}$ is the number of snapshots, $M_{200}(z_i)$ is the virial mass at the redshift of the $i$th snapshot, $z_i$, for the halo from the grid (or CCVT) simulation, and $M_{\rm 200, glass}(z_i)$ is the virial mass of the matched halo at $z_i$ in the glass simulation.

\begin{equation}
    \Delta_{N_{\rm s}}(x) \equiv \frac{1}{N_{\rm s}}\sum_{i=1}^{N_{\rm s}} \left|x(z_i) - x_{\rm glass}(z_i) \right|/x_{\rm glass}(z_i)
\end{equation}
quantifies the averaged relative difference of halo property $x$ over $N_{\rm s}$ snapshots. $x(z_i)$ is the property of a halo from the grid (or CCVT) simulation at $z_i$, while $x_{\rm glass}(z_i)$ is the property of a halo from the glass simulation at $z_i$. $x$ stands for any of the halo properties discussed in this work. 

Note that both $Q_{N_{\rm s}}$ and $\Delta_{N_{\rm s}}(x)$ are dimensionless. The smaller $Q_{N_{\rm s}}$ [$\Delta_{N_{\rm s}}(x)$] is, the better the halo mass accretion history (the evolution of halo property $x$) converges. 

In the following, we set $N_{\rm s} = 20$ and use the last 20 simulation snapshots, which corresponds to the evolution history of a halo in the last $\sim 2$ Gyr (or from $z \sim 0.15$ to $z=0$) , to compute $Q_{20}$ and $\Delta_{20}(x)$ for all haloes in the MATCHED sample. The correlations between $Q_{20}$ and $\Delta_{20}(x)$ are shown in Fig. \ref{fig:Q20_D20_correlation} for seven properties (i.e. except for the halo formation time which is specifically defined for $z=0$ haloes). We have also computed the Spearman's rank coefficients and the corresponding p-values to quantify the correlation strength for each halo property (shown in the lower part of each panel).

We can find that for all halo properties, there are weak or moderate positive correlations between $\Delta_{20}(x)$ and $Q_{20}$, indicating that the poor numerical convergence in halo properties tends to associate with largely different mass growth histories. This confirms the aforementioned observations from individual haloes.

Note that we have also tested with $N_{\rm s} = 5$, $N_{\rm s} = 10$ and $N_{\rm s} = 30$, and confirmed that our conclusions above do not change. We have also looked at the evolution of halo properties for outliers in the matched halo sample from the glass-1 versus glass simulations as well as from the CCVT-1 versus CCVT simulations, and the results are similar to what have been discussed above for the MATCHED outliers from the grid versus glass and CCVT versus glass simulations.

\bibliographystyle{mnras}
\bibliography{paper}         

\bsp	
\label{lastpage}

\end{document}